# Modeling Network Security: Case Study of Email System

Sabah Al-Fedaghi[1]
Computer Engineering Department
Kuwait University
Kuwait

Hadeel Alnasser[2]
Technical Support Department –
Kuwait Anti-Corruption Authority
Kuwait

*Abstract*—We study operational security in computer network security, including infrastructure, internal processes, resources, information, and physical environment. Current works on developing a security framework focus on a security ontology that contributes to applying common vocabulary, but such an approach does not assist in constructing a foundation for a holistic security methodology. We focus on defining the bounds and creating a representation of a security system by developing a diagrammatic representation (i.e. a model) as a means to describe computer network processes. The model, referred to a thinging machine, is a first step toward developing a security strategy and plan. The general aim is to demonstrate that the *representation* of the security system plays a key role in making thinking visible through conceptual description of the operational environment, a region in which active security operations are undertaken. We apply the proposed model for email security by conceptually describing a real email system.

*Keywords*—*Network security; conceptual model; diagrammatic representation; email system*

## I. Introduction

Security typically refers to the state of being secure; that is, being free from danger (the reciprocal of safety). A security system involves deterrence and prevention of threats and is internally motivated as a self-protecting system. It is possible to be in a safe state while danger is present. We can represent this with the notion of threat as a point on a scale between safety and harm that indicates potential danger. Threat also indicates the degree of preparedness required to achieve security.

Preparedness contrasts with vulnerability. Given a certain degree of threat, security is a measure of preparedness for confronting such a threat. The first requirement of preparedness is defining the bounds and creating a representation of the operational environment. The first phase in operational security involves identifying critical assets (e.g., data) and an infrastructure in the operational environment [1]. From the military's perspective [2], the first step in characterizing the operational environment involves identifying the operational area and determining the significant characteristics and level of detail.

According to Tolone et al. [2], "The success of defense and security operations depends on the ability to make sense of the operational environment and to anticipate those factors that influence operations both negatively and positively." Making sense refers to creating situational awareness and employing continuous effort to understand connections, anticipate their trajectories, and act effectively [2].

In this context, this paper imports the sense of operational security in computer network security, including infrastructure, internal processes, resources, information, and physical environment. We focus on defining the bounds and creating a representation by developing a diagrammatic representation (i.e. model) as a means to describe computer network facilities. It is a first step toward developing a security strategy and plan.

### A. The Security Problem

Researchers have categorized some security problems as "wicked problems" due to their complexity, intricacy, and intractability [3]. According to Gilmore [4], "Many cyber issues personify a wicked problem. Cyber security can never truly be solved (a completely secure network is a myth)."

Huguet [5] examined the security notion in general, concluding, "Nowadays there are several 'securities' and a number of models from different fields. Despite the importance of the issue, surprisingly, there is no common vocabulary, procedures, definition or model to share knowledge about security." Having a general framework of the security concept, in which to integrate those models and concepts, has advantages, such as shared vocabulary, knowledge, development, or metrics [6-7]. Solms and Solms [7] emphasized that "with the need to implement IT-security measures in almost every environment. Holistic security ontology is still missing. We have to model proper countermeasures capable of protecting the resources." The authors also stressed infrastructure elements such as electronic devices and networks, as well as their relationships.

We note that current works on developing a security framework focus on a security ontology that contributes to applying a common vocabulary, but such an approach does not assist in constructing a foundation for a holistic security methodology: "a holistic formal graphical and textual paradigm for the representation, development, and lifecycle support of complex systems" [8]. Moreover, in security, "the problem lies in the details" [9]. Maintaining security networks with heterogeneous systems, policies, and capabilities quickly became a major task because system administrators were required to maintain detailed descriptions of each host [9].

### B. Contributing to the Solution

This paper focuses on developing a modeling language that can be utilized to build a security foundation. The objective is to demonstrate that the representation of the security system plays a key role in making thinking visible, through conceptual description of the security of the operational environment, a region in which active security operations are undertaken. Fig. 1 and 2 show two descriptions of such a theater of operations.

Our proposed representation (model) is constructed in terms of a diagram that serves at the level between "natural communication" and semiformal specification to facilitate understanding among all security participants as a first step toward developing and implementing security policies and implementation plans.

### C. Focus

Without loss of generality, in this paper we focus on developing a foundation for email security through conceptually describing what we previously called a theater of operation involving email. Today, email has become the backbone of many professionals' daily activity. Emails are most frequently used in commerce [12]. In everyday life, we rely on email's confidentiality and integrity to exchange data and communication.

According to Landewe [13], email is the primary threat to companies using enterprise platforms, such as Office 365. Email security aims to develop an email technology with a more innovative and multilayered approach to cloud security.

New email security technology involves monitoring user behavior and events, as well as greater access to files, users, and controls. This allows suspicious email to be caught before it reaches an inbox. Using API, it is instead held in a quarantine folder. A copy of the email is run through various technologies (e.g., sandboxing) [13]. The identification of suspicious email is accomplished by performing language and contextual analysis and business email compromise and phishing analysis. Emails with a URL must be handled with link analysis using real-time feeds or by sandboxing the URL [13].

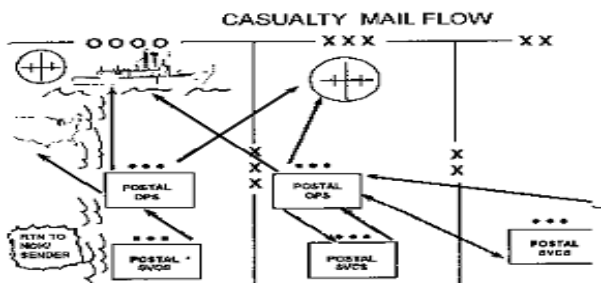

Fig. 1.   Mail Theater of Operation (adapted from [10]).

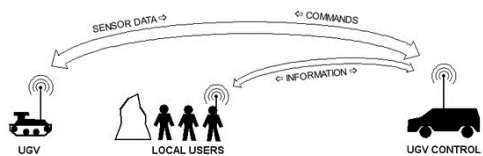

Fig. 2.   Sample Description of a Theater of Operation (adapted from [11]).

## II.   Related Works

The email system architecture is typically introduced when discussing email issues. Two samples of such representations are shown in Fig. 3 and 4. Many types of email network diagrams exist, and UML diagrams are also used (e.g., use-case diagram). Fig. 5 shows a sample of email models. Other works use architectural diagrams that supplement a hardware-oriented network to investigate logical data flow embedded in a system. This approach arrives at a variation of the UML diagrams and includes actual hardware connectivity and logical flow of data (e.g. [14]). In UML, the multiplicity of diagrams is a known problem [8] when what is needed is a single, integrated diagrammatic representation that incorporates function, structure, and behavior.

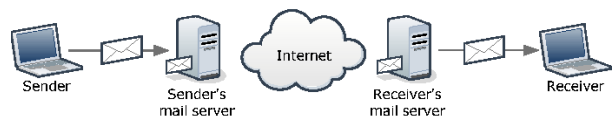

Fig. 3.   Email System Architecture (Partially Adapted from [15]).

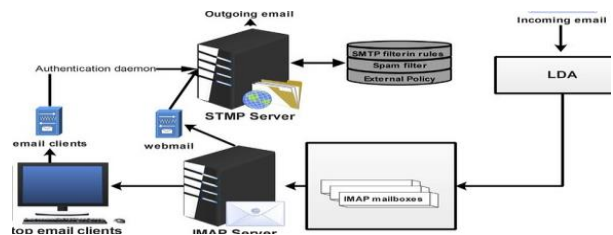

Fig. 4.   Email System Architecture (Partially Adapted from [16]).

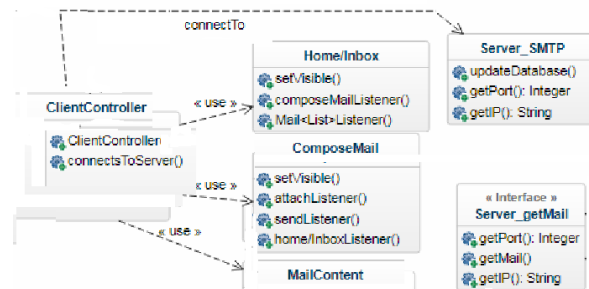

Fig. 5.   Sample UML Email System Model (Partially Adapted from [17]).

## III.   Thinging Machine

According to Wong [18], tackling wicked problems can be achieved by conceptualizing them as systems and breaking them down into "chunks of information" or digestible nodes and their relationships. In describing the model for email operational security, we use the thinging machine (TM) model [19-28] where all operational elements are conceptualized in terms of a single ontological entity, the thimac (thing/machine). As we show, thimacs can represent heterogeneous entities: physical entities (e.g., a server, router, or workstation), software objects (e.g., a program or software system), and other notions (e.g., protocols, flows, or plans).

The term thing (in contrast to objects in object-oriented modeling) indicates an expansive specification of an entity that reflects one mode of an entity's being. Heidegger's [29] notion was that a thing is not a mere abstract object but something that is operated on (created, changed, and transported) and, simultaneously, a process (machine) that subjects other things to its activities (creating, changing, and transporting). According to Heidegger [29], a thing "things"; that is, it gathers, unites, or ties together its constituents, in the same way that a bridge unifies aspects of its environment (e.g., a stream, its banks, and the surrounding landscape) [30].

Building on such a philosophical approach, things are combined with process by viewing them as blocks of single ontological thimacs, which populate a world that is also a thimac (we call it a system). In contrast to the object-oriented paradigm, every part of this world is a thimac, forming a thimac-ing network. A unit of such a universe has dual being as a thing and as a machine. A thing is created, processed, released, transferred, and/or received. A machine creates, processes, releases, transfers, and/or receives things. We will alternate between the terms thimac, thing, and machine according to the context.

The thimac as a {thing, machine} pair designates what simultaneously divides and brings together a thing and a machine (process in the general sense). Every thimac appears in a system either by creation or importation from outside the system. They are the concomitants (required components) of a system and form the current fixity of being for any system that continuously changes from one form (thing/machine) to another. We will use the notion of thimac to model an email system focusing on security aspects.

The terms system and model have been used ubiquitously in engineering [31]. In TM, a system is the overall constellation of thimacs that structures all subthimacs in the problem under consideration. It provides the problem's unifying element through space and time as integral subthimacs, not as the sum of individual subthimacs. Thimacs inside a system are understood not as things with properties but as ensembles of things and machines that constantly interact with each other and with the out-of-system world.

In this complex model, events (a type of thimac that involves time) appear, propagate, and constantly recur in various parts of the system with repeatable occurrences and stable regularities. In its static and dynamic modes, the whole system is a representation (mimesis) of a portion of reality.

Accordingly, a thimac's existence depends on its position in the larger system, as either a thing that flows in other machines or a machine that handles a flow of things (i.e., create, process, release, transfer, and receive things). It brings together and embraces both "thingishness" and "machineness." A thing's flow is conceptualized as an abstract structure that forms an abstract machine called a TM (Fig. 6), in which the elementary processes are called the stages of a TM. In the TM model, we claim that five generic processes of things exist: things can be created, processed, released, transferred, and received. These five processes form the foundation for modeling thimacs. Among the five stages, flow (solid arrow in Fig. 6) signifies conceptual movement from one machine to another or among the stages of a machine. The TM stages can be described as follows.

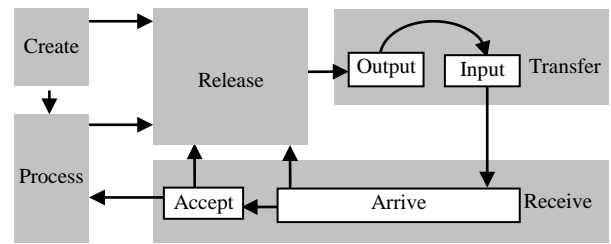

Fig. 6. A Thinging Machine.

Arrival: A thing reaches a new machine.

Acceptance: A thing is permitted to enter the machine. If arriving things are always accepted, then arrival and acceptance can be combined into the receive stage.

Processing (change): A thing undergoes some kind of transformation, without creating a new thing.

Release: A thing is marked as ready to be transferred outside of the machine.

Transference: A thing is transported somewhere outside of the machine.

Creation: A new thing is born (created) in a machine. "Create" resembles "there is."

In addition, the TM model includes memory and triggering (represented as dashed arrows) relations among the processes' stages (machines).

## IV. EXAMPLE OF A THINGING MACHINE

To illustrate TM modeling, in this section we model a communication protocol in widespread use today. One definition of a protocol is a standard description used to define a method of exchanging data over a computer network [32]. In TM, a protocol is viewed as a thimac (machine) that is formed from subthimacs that create, process, release, transfer, and/or receive things (e.g., signals, data, or messages). The protocol is also a thing that can be created and processed (e.g., updated). We apply TM modeling to the well-known simple mail transfer protocol (SMTP).

SMTP is an Internet standard for email transmission [33-34]. It allows for a simple email service and is responsible for moving messages from one email server to another. SMTP includes many standard commands (e.g., EHLO, MAIL FROM, RCPT TO, DATA, and QUIT).

The SMTP protocol and how email works can be explained by tracking the journey of an email message from one person, say, Bob, to another, Alice. Bob composes his message and after inserting Alice's email address, he clicks the "send" button. SMTP governs the communication between Bob's mail server and Alice's mail server [35]. Fig. 7 shows a sequence diagram that models all the events (in the UML sense) involved in this communication, assuming everything works correctly. This diagram is shown in many sources on the Internet (e.g., [35][36][37]).

Fig. 8 shows the corresponding TM model, in which the whole protocol is constructed from the two subthimacs: Bob's mail server and Alice's mail server. In the first machine, the EHLO message (identifying domain, e.g., Gmail.com) is constructed and sent by Bob's mail server (Circle 1 in Fig. 8). The message flows to Alice's mail server, where it is processed (2) to create a response message (3) of acknowledgement (4), along with the name of the email services that the SMTP server can support (e.g., Yahoo.com). Bob's mail server sends the sender's email address (5; e.g., Bob@gmail.com), which flows to Alice's mail server to trigger an OK message (6) that reaches Bob's mail server. Afterwards, Bob's mail server sends the email address of the recipient (e.g., Alice@yahoo.com) (7) to trigger Alice's mail server to reply with an OK message (8).

At this point, Bob's mail server requests that the data part of the email be sent (9), and upon receiving a ready message from Alice's mail server (10), Bob's mail server starts sending the data (11) line by line. Upon sending the whole message, Alice's mail server sends an acceptance of the message (12). Upon receiving the message, Bob's mail server requests to quit (13) and Alice's mail server sends a signal to close the connection (15), which is closed as the last step (15).

In contrast to the sequence diagram in Fig. 7, the representation in Fig. 8 is based on TM. Although the sequence diagram potentially includes millions of arbitrary actions (e.g., send, identify, terminate, receive, respond, accept, and close), the TM specification repeatedly uses five generic operations. Even though Fig. 8 has the appearance of a complex structure, this complexity is a visual impression that emerges from this repeated application of TM. According to Bishop [38], systems that have a complicated set of interacting parts may actually exhibit relatively simple behavior.

Fig. 8 models the static description of SMTP. To model the dynamic behavior, we use events. An event in TM is a thimac that includes a time machine. For example, Fig. 9 shows the event Sending a line of data. Because a hierarch of events exists in the SMTP example, we select the 12 events in Fig. 10. Accordingly, Fig. 11 shows the dynamic behavior of the SMTP system.

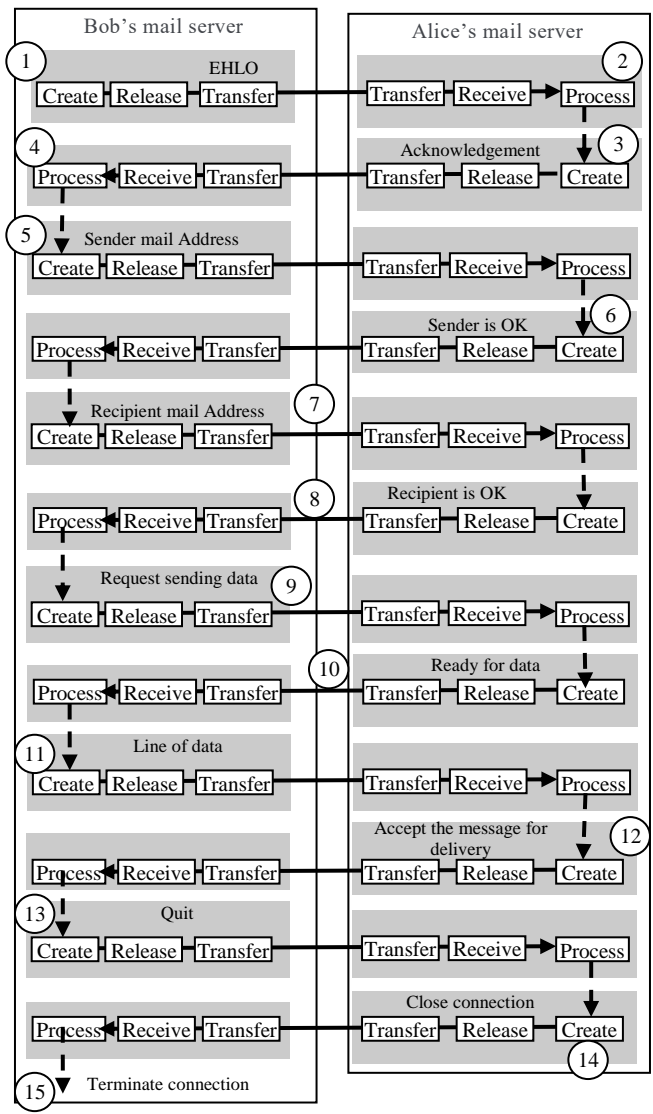

Fig. 8. The SMTP Thinging Machine.

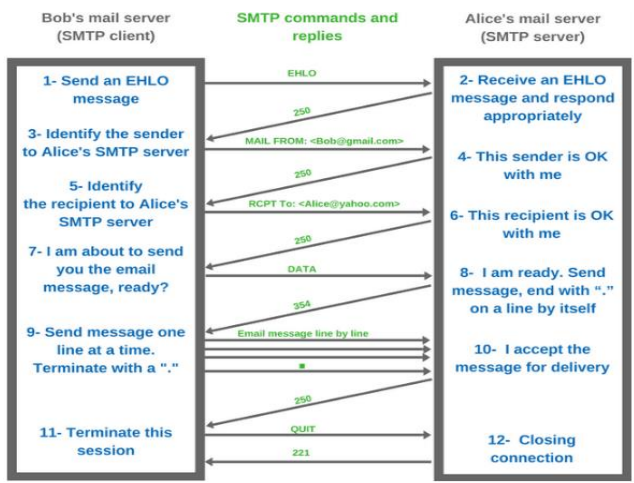

Fig. 7. Representation of SMTP as a Sequence Diagram.

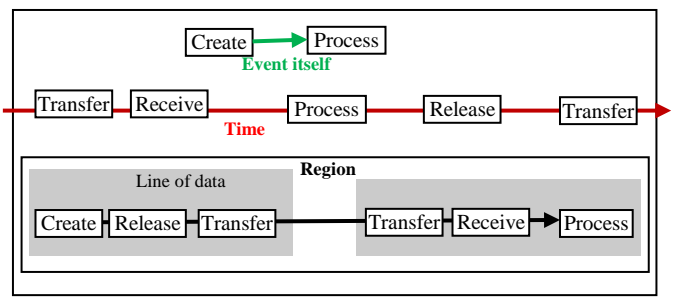

Fig. 9. The Event Sending a Line of Data.

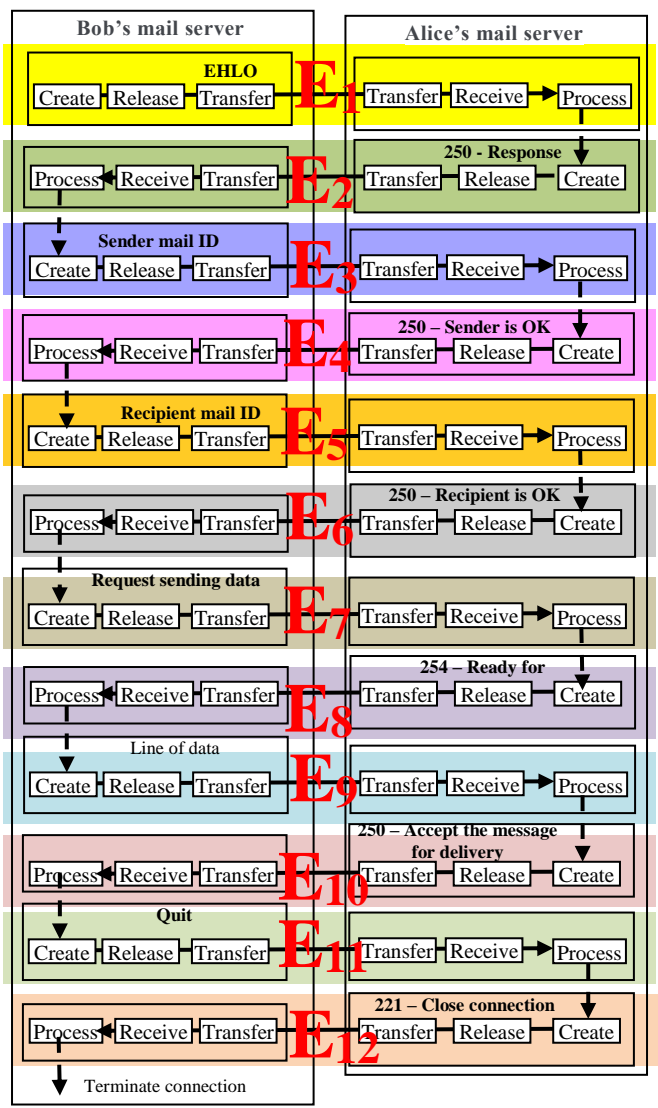

Fig. 10. Events of the SMTP Thinging Machine.

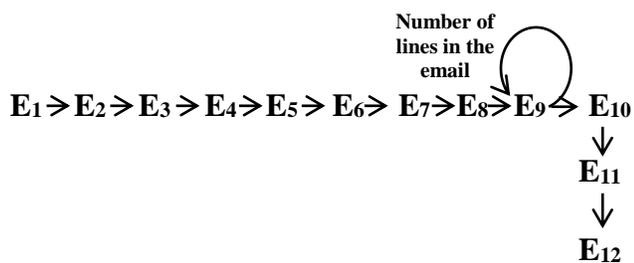

Fig. 11. The Chronology of Events in the SMTP Thinging Machine.

## V. CASE STUDY

The map of the email system's security control is essential for understanding, implementation, expansion of the design, training, documentation, and management. The map includes descriptions of each email security device and system that participates in the network. In this case study, we develop such a map for a currently existing government network (workplace of the second author). We limit our mapping to tracking the email throughout the network. This involves diagramming the creation and processing stages of the email's flow from the user workstation to its destination. This includes the following components.

*1)* User's workstation (e.g., any smart device that can access the organization's email system).

*2)* Email system: The email system that includes servers that facilitate emailing across the network.

*3)* Internal firewall: An internal firewall only allows legitimate traffic based on configured policy and rules.

*4)* Email security gateway: An email security gateway prevents the transmission of emails that violate policy, malware, or transfer of information with malicious intent.

*5)* External firewall: An external firewall only connects an internal network to an external network and all services or published servers, along with third-party connection (in our case Internet), to separate and secure internal networks and traffic.

*6)* Domain Name System (DNS) server: A DNS server is a computer server that contains a database of public IP addresses and their associated host names, and in most cases, serves to resolve or translate those names to IP addresses as requested (e.g., www.google.com will be translate to 8.8.8.8). DNS servers run special software and communicate with each other using special protocols.

*7)* Internet service provider router: A router provides access to the Internet and transfers the traffic from the external firewall to the Internet cloud.

Fig. 12 shows a general picture of the connections among these components of our case study.

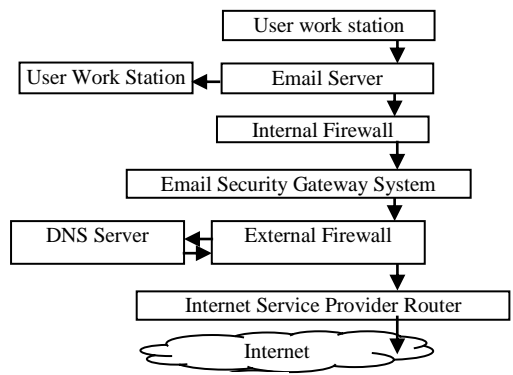

Fig. 12. A General View of the Modelled Network.

## VI. TM EMAIL MODEL

Fig. 13 shows the TM model of the system.

### A. In the user Workstation

In the figure, the email process starts when the user, on his or her workstation (1), creates an email (2) using the email system (3). This process involves the following.

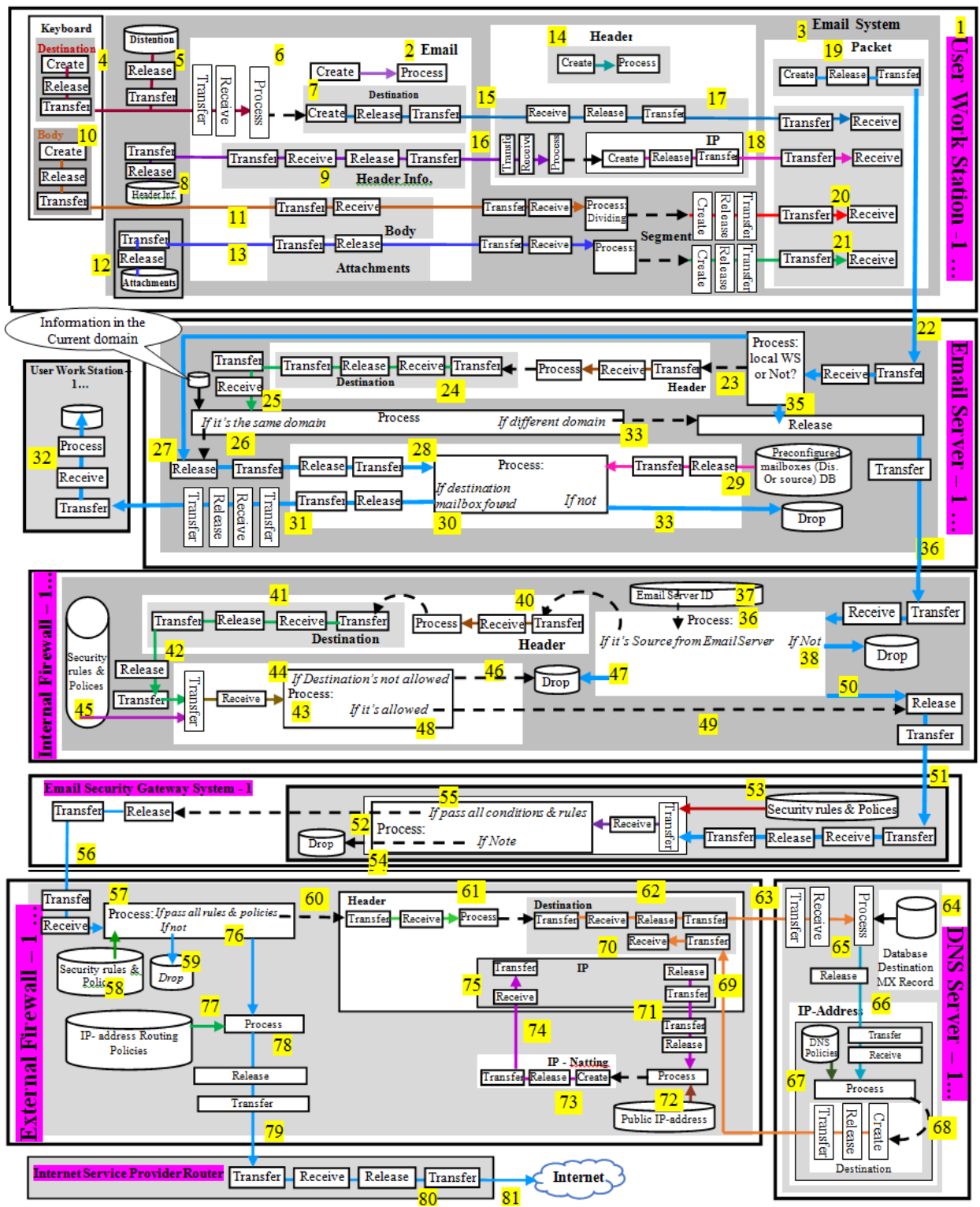

Fig. 13. The TM Model of the Email System.

*1)* The destination address is either typed in using the keyboard (4) or retrieved from storage (5) to be processed (6) in the email system to create an initial destination format (7).

*2)* In addition, the header format (8) is retrieved and flows to the email (9).

*3)* The email body is created (10) and flows to the email (11).

*4)* If there is attachment (12), then it flows to the email (13).

*5)* The header is created (14) by combining the destination (15) and initial header information (16) to fill the IP address fields.

*6)* The email packet is generated (19) by conjoining the header IP address field information (17, 18), body (20), and attachments (21).

### B. In the Email Server

The email packet leaves the user's workstation (22) and reaches the email server (known as the exchange server). The packet is processed in the email server to extract the header (23), and then the destination is extracted (24) for comparison with the information in the current domain.

*1)* If the destination has the same current information domain (26), then the packet flows (27) to be processed (28) for comparison with preconfigured mailboxes (29).

- If a destination mail box is found (30), then the packet flows (31) to the user's work station (32).
- If the destination mail box is not found (33), then the packet is dropped.

*2)* If the destination does not have the same current information domain (34), then the packet flows (35) to the internal firewall (36).

### C. In the Internal Firewall

In the internal firewall, the packet is checked (37) for an email server ID (38):

*1)* If its source is not from the current email server (39), then it is dropped (40).

*2)* If its source is from the current email server, then the header is extracted (41) and the destination in the header is extracted (42). Then the destination flows (43) to be processed (44).

-If the destination is not permitted (45) by the security rules and polices (46), then the packet is dropped (47, 48).

-If the destination is permitted (49) by the security rules and polices (46), then the packet flows to the email security gateway system (50, 51, and 52).

### D. In the Email Security Gateway System

In the email security gateway system, the packet is compared (53) with the security rules and polices (54).

*1)* (i) If the packet does not satisfy all polices and rules, then it is dropped (55).

*2)* (ii) If the packet is passes all polices and rules (56) then it's flow to the External Firewall (56).

### E. In the External Firewall

The packet is processed (57) in the external firewall.

*1)* If the packet does not satisfy the security rules and polices (58), then it is dropped (59).

*2)* If the packet satisfies all polices and rules (60), then the header information (61) and the destination (62) are extracted from the header. The destination flows to the DNS server (63) to be compared with the stored DNS database records (64) to select the related destination MX record (65). The MX record, known as the mail exchanger record, specifies the mail server responsible for accepting email messages on behalf of a domain name. It is a resource recorded in the DNS. This record flows to be processed again (66 and 67) with the DNS polices to create the destination IP address (68) that flows to the header (69 and 70).

The IP address of the header flows to be processed (71) with the public IP address (72) to be processed again to create the natted public IP address that is used for the routing polices (73), which flows again to the header (74 and 75).

Once the header is updated, the packet is released (76) to be processed (77) with the IP address routing polices (78) to learn its next destination, then it travels (79) to the Internet service provider router.

### F. In the Internet Service Provider Router

The email packet leaves the external firewall and is transferred (80) to the cloud (81).

## VII. TM DYNAMIC MODEL

Note that the static email model structure is formed from the flows among thimacs in the system. It also includes the network of thimacs belonging to the system. The resultant conceptual model is a representation of structure in the email system. An email has a class (e.g., object-oriented) form with header, data, and attachment attributes (object-oriented terminology). These are filled by flows to produce an object (object-oriented terminology). The behavior is yet to be defined when we incorporate events in the static model. Note the two forms of a thimac. For example, the packet is a machine that is fed addresses, pieces of data, and attachments. As soon as it is loaded, it flows as a thing to the email server.

A thimac is activated by elevating it to a time thimac (e.g., a subdiagram of Fig. 13 becomes the region of an event). We can develop the dynamic model of the email system as we did before, for the SMTP protocol. However, in consideration of space, we only identify events in the user workstation and the pre-email server as shown in Fig. 14. Thus, what appeared in Fig. 13 as the transmission of one packet, in the dynamic model with its chronology of events (Fig. 15), we see as a repeated generation of packets until all data and attachments have been sent.

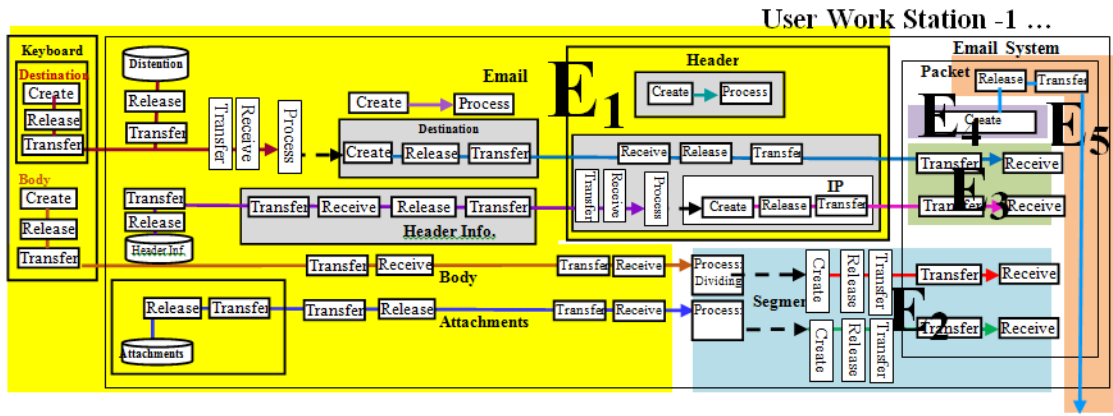

Fig. 14. Selected Events in the user Workstation.

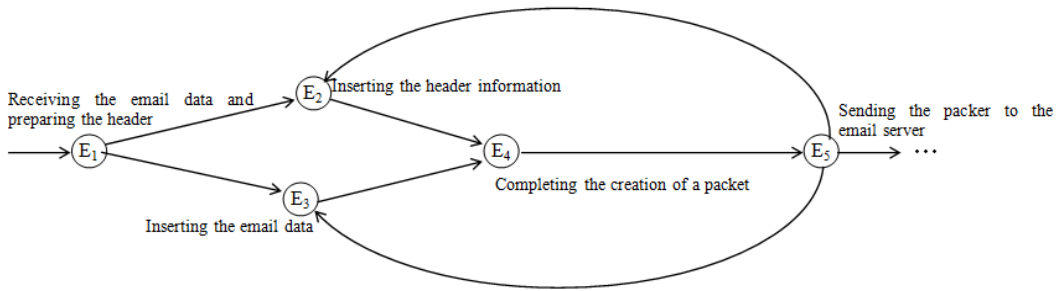

Fig. 15. Chronology of Events in the user Workstation.

## VIII. TM SECURITY MODELING

In the TM network model, we can separate separate the network security system from the functional system. Each security thimac (hardware and software) can be described by tracking the packet flow. The ad hoc diagrams and flowcharts currently in use (e.g. see Fig. 16 and 17) are "crude" representations that are not as systematic as the TM model, which involves only five primitive operations. In addition, these current representations do not model dynamic aspects in the same diagram, e.g., events of firewall-1 as shown in Fig. 18.

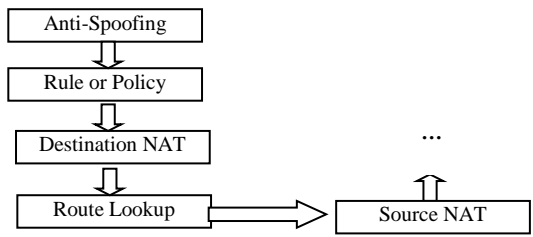

Fig. 16. Packet flow Check Point Firewall (Partially Adapted from [39]).

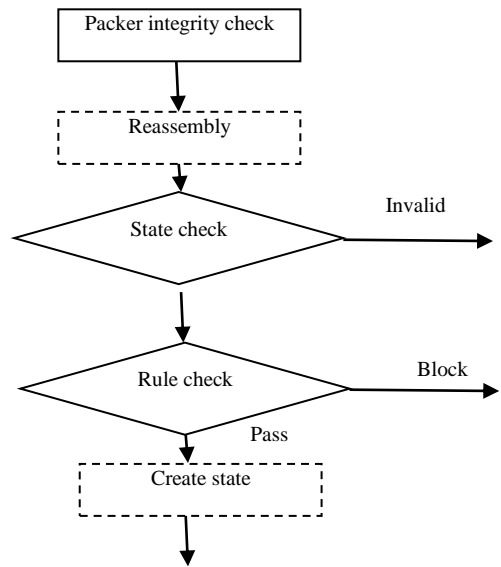

Fig. 17. Packet Processing in Firewall (Partially Adapted from [40]).

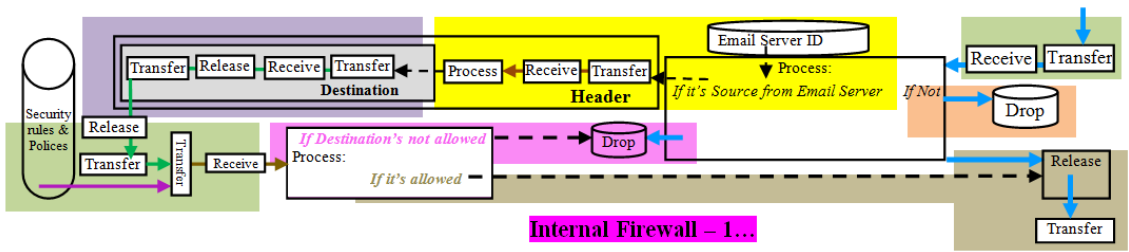

Fig. 18. Events in Firewall-1 (see Fig. 13).

## IX. CONCLUSION

In this paper, we claim that the task of building a security system requires construction of a diagrammatic representation of the operational environment where protected assets reside. We criticized current modeling languages (e.g., UML and ad hoc diagrams) as languages that either include large heterogeneous notions or nonsystematic symbols (wall, cloud, screen, human figure, etc.). Instead, we have proposed using TM modeling and applied TM language to an actual email system by tracking security aspects as an email flows through various system components. The resultant map can be used as a foundation for the activities of an email security officer, just as a network diagram is a tool for a network engineer. Future work will apply the TM model to different types of networks.

### REFERENCES

[1] J. Andress, "Operations security," in The Basics of Information Security (2nd ed.). Elsevier, 2014.

[2] W. J. Tolone, X.. Wang and W. Ribarsky, Making Sense of the Operational Environment through Interactive, Exploratory Visual Analysis, NATO OTAN unclassified paper.

[3] P. Williams, Security Challenges as Wicked Problems, Stratfor, Jun 1, 2018.

[4] K. Gilmore, "Cyber security: A wicked problem," Air Force Information Technology & Cyberpower Conference, Aug 29-31, 2016, Montgomery, AL, 2016.

[5] M. Colobran Huguet, "A general-purpose security framework," Ph.D. thesis, Universitat Autònoma De Barcelona, September 2015.

[6] S. Fenz and E. Weippl. "Ontology based IT-security planning," in Proceedings of the 12th Pacific Rim International Symposium on Dependable Computing, pp. 389–390. IEEE Computer Society, University of California, Riverside, 2006.

[7] B. von Solms and R. von Solms, "From information security to business security?" Computers & Security, Vol. 24, No. 4, pp. 271–273, 2005.

[8] D. Dori, Object-Process Methodology - A Holistic Systems Paradigm. Berlin: Springer Verlag, 2002.

[9] M. Carvalho, M. Rebeschini, J. Horsley, N. Suri, T. Cowin, M. Breedy, "MAST: Intelligent roaming guards for network and host security," Scientia, Estudos Interdisciplinares em Computação, Vol. 16, No. 2, pp. 125-138, December 2005.

[10] GlobalSecurity.com, Postal operations management, Chapter 6, 2020.

[11] H. Sallum, "Development and implementation of a high-level command system and compact user interface for non-holonomic robots," M. S. thesis, Massachusetts Institute of Technology, May 2005.

[12] S. Choudhary, "E-mail security: Issues and solutions," International Journal of Computer Information Systems, Vol. 7, No.4, 42-46, 2013.

[13] J. Witts, "M. Landewe interview," Expert Insights, Nov 20, 2019.

[14] T. Olzak, "A practical approach to threat modeling," Erudio Security, LLC, 2006.

[15] A. Z. Adamov, "Internet technologies in depth. The technique of spam recognition based on header investigating," 5th International Conference on Application of Information and Communication Technologies (AICT),1 - 5, Azerbaijan, Baku, 12-14 October 2011.

[16] E. Gbenga Dada, J. S. Bassi, H. Chiroma, S. M. Abdulhamid, A. O. Adetunmbi, and O. E. Ajibuwa, "Machine learning for email spam filtering: Review, approaches and open research problems," Heliyon Vol. 5, No. 6, 2019.

[17] Oshinsingh, "Email system, UML model," GenMyModel, September 24, 2014.

[18] E. Wong, "Wicked problems: 5 steps to help you tackle wicked problems by combining systems thinking with agile methodology," The Interaction Design Foundation, 2019.

[19] S. Al-Fedaghi and A. J. Al-Fadhli, "Thinging-oriented modeling of unmanned aerial vehicles," International Journal of Advanced Computer Science and Applications (IJACSA), Vol. 11, November, 2019.

[20] S. Al-Fedaghi and Y. Atiyah, "Tracking systems as thinging machine: A case study of a service company," International Journal of Advanced Computer Science and Applications (IJACSA), Vol. 9, No. 10, pp. 110-119, 2018.

[21] S. Al-Fedaghi and M. BehBehani, "Thinging machine applied to information leakage," International Journal of Advanced Computer Science and Applications (IJACSA), Vol. 9, No. 9, pp. 101-110, 2018.

[22] S. Al-Fedaghi and M. Al-Otaibi, "Conceptual modeling of a procurement process: Case study of RFP for public key infrastructure," International Journal of Advanced Computer Science and Applications (IJACSA) – Vol. 9, No. 1, January 2018.

[23] S. Al-Fedaghi and N. Al-Huwais, "Conceptual modeling of inventory management processes as a thinging machine," International Journal of Advanced Computer Science and Applications (IJACSA), Vol. 9, No. 11, pp. 434-443, November 2018.

[24] S. Al-Fedaghi, "Thinging as a way of modeling in poiesis: Applications in software engineering," International Journal of Computer Science and Information Security (IJCSIS), Vol. 17, No. 11, November 2019.

[25] S. Al-Fedaghi, "Thing/Machine-s (thimacs) applied to structural description in software engineering," International Journal of Computer Science and Information Security (IJCSIS), Vol. 17, No. 8, August 2019.

[26] S. Al-Fedaghi, "Five generic processes for behaviour description in software engineering," International Journal of Computer Science and Information Security (IJCSIS), Vol. 17, No. 7, pp. 120-131, July 2019.

[27] S. Al-Fedaghi, "Modeling events and events of events in software engineering," (IJCSIS) International Journal of Computer Science and Information Security, Vol. 18, No. 1, 2020.

[28] S. Al-Fedaghi, "Thing/machine-s (thimacs) applied to structural description in software engineering," International Journal of Computer Science and Information Security (IJCSIS), Vol. 17, No. 8, August 2019.

[29] M. Heidegger, "The thing," in Poetry, Language, Thought, A. Hofstadter, Trans. New York: Harper & Row, 1975, pp. 161–184.

[30] M. Heidegger, Being and Time, J. Macquarrie and E. Robinson, Trans. London: SCM Press, 1962.

[31] D. Hestenes, "Notes for a modeling theory of science, cognition and instruction," in E. van den Berg, A. Ellermeijer & O. Slooten, Eds., Modelling in Physics and Physics Education. U. Amsterdam, 2008.

[32] N. Emberton, "Protocol," Computer Hope, Oct. 7, 2019.

[33] J. B. Postel, "Simple mail transfer protocol," Information Sciences Institute, University of Southern California, August 1982.

[34] J. Klensin, "Simple mail transfer protocol," Network Working Group, October 2008.

[35] K. Elghamrawy, "SMTP protocol explained (How email works)," Afternerd Blog, 2017-2019.

[36] C. M. Kozierok, The TCP/IP Guide, September 20, 2005.

[37] TeleMessage, "How to know if the mail server on the other side received an email," TeleMessage Site, 1999-2020.

[38] R. C. Bishop, "Metaphysical and epistemological issues in complex systems," in Handbook of the of Science Philosophy, Vol. 10, edited by Cliff Hooker, North Holland, 2011.

[39] F. Ali, "Checkpoint firewall packet flow," Network Engineer, April 28, 2019.

[40] Oracl Documentation Center, Securing the network in Oracle Solaris 11.4, January 2019.